\documentclass[aps,pre,epsfigure,twocolumn,superscriptaddress]{revtex4-1}
\usepackage[colorlinks=true,linkcolor=blue,urlcolor=blue,citecolor=blue,pdfusetitle]{hyperref}
\usepackage[utf8]{inputenc}
\usepackage[english]{babel}
\usepackage{amsmath}
\usepackage[caption = false]{subfig}
\usepackage{graphicx,epstopdf}
\usepackage{blindtext}
\usepackage[table,xcdraw]{xcolor}
\usepackage{lipsum}
\usepackage{amsfonts}
\usepackage{bbm}
\usepackage{amssymb}
\usepackage{enumerate}
\usepackage{color}
\usepackage{latexsym}
\usepackage{physics}
\usepackage{times,txfonts}

\begin{document}
	
\title{Environment-mediated charging process of quantum batteries}

\author{F. T. Tabesh}
\email{f.tabesh@uok.ac.ir}
\affiliation{Department of Physics, University of Kurdistan, P.O.Box 66177-15175 , Sanandaj, Iran}

\author{F. H. Kamin}
\affiliation{Department of Physics, University of Kurdistan, P.O.Box 66177-15175 , Sanandaj, Iran}

\author{S. Salimi}
\email{ShSalimi@uok.ac.ir}
\affiliation{Department of Physics, University of Kurdistan, P.O.Box 66177-15175 , Sanandaj, Iran}

\date{\today}

\begin{abstract} 
We study the charging process of open quantum batteries mediated by a common dissipative environment in two different scenarios. In the first case, we consider a quantum charger-battery model in the presence of a non-Markovian environment. Where the battery can be properly charged in a strong coupling regime, without any external power and any direct interaction with the charger, i.e., a wireless-like charging happens. The environment plays a major role in the charging of the battery, while this does not happen in a weak coupling regime. In the second scenario, we show the effect of individual and collective spontaneous emission rates on the charging process of quantum batteries by considering a two-qubit system in the presence of Markovian dynamics. Our results demonstrate that open batteries can be satisfactorily charged in Markovian dynamics by employing an underdamped regime and/or strong external fields. We also present a robust battery by taking into account subradiant states and an intermediate regime. Moreover, we propose an experimental setup to explore the ergotropy in the first scenario.
\end{abstract}
\maketitle
\section{Introduction}
Every quantum system can be considered as an open system because of the unavoidable mutual interaction with an environment. The interaction between the system and the environment generally results in the loss of typical quantum properties such as coherence or entanglement as well as energy dissipation \cite{25}. On the other hand, the preservation of such quantum properties to better energy storage is a fundamental topic.  Therefore, one cannot ignore the dissipative effects of the environment on the stable charging of quantum batteries ($QB$s). This important issue leads to the study of open quantum batteries \cite{13,Ba,14,15,16,28,19}.

$QB$s are $d$-dimensional quantum devices with non-degenerate energy levels that are used to temporarily store energy in quantum degrees of freedom, in order to transfer energy from production to consumption  hubs \cite{1,3}.  Operationally, an optimal $QB$ needs to have two important factors: First, the maximum average charging power (the maximum stored energy with the minimum charging time). Second, the capability to fully transfer the stored energy to consumption centers in a useful way (the skill of  extracting useful work)\cite{2,4,5,6,7,8, e1,e2}. Therefore, providing protocols to accomplish these two aims are particularly significant.
 
 In a very new view, the $QB$s have been considered as open systems, where the battery, the charger, or both are in contact with a reservoir \cite{13,Ba,14,15,16,28,19}. Although
many studies have been done on environmental effects on charging process, but less effort has been devoted to the role of the memory effects and spontaneous emissions on battery efficiency in a common reservoir\cite{Ba,14,15,16,28,19}. Given the importance of the topic and the availability of some efforts that cover various aspects of the physics of open quantum batteries, we will focus here on a specific issue and that is the role of common reservoirs with non-Markovian and Markovian dynamics \cite{25,29} to providing an optimal protocol compared to others. Here, unlike the mechanism presented in Ref.~\cite{13}, in addition to a charge-mediated energy transfer for the open quantum batteries, we are also dealing with an environment-mediated case. In fact, our model allows energy to leak from the battery to the environment and realize a realistic scenario of spontaneous discharge of quantum batteries. Where we take advantage of both the memory effects induced by non-Markovian dynamics and the diversity of the coupling regimes to prevent such a phenomenon. We also discuss the model presented in Ref.~\cite{15} with a different perspective by regarding non-Markovian environment and spontaneous emissions. 
 
In this paper, we investigate the process of charging in two scenarios. First, we consider a charger-battery model via a two-qubit system. In which the battery in the absence of external fields and a direct connection with the charger can be charged under non-Markovian dynamics or in the so-called wireless-like charging process. In this case, the battery is satisfactorily charged by the mediation of the environment in the strong coupling regime, while this is not observed in the weak coupling regime. In the second scenario, we consider the destructive effects of spontaneous emissions on the charging process by regarding a two-qubit system in the presence of Markovian dynamics, such that there is a dipole-dipole interaction between the qubits and each qubit can be charged through a laser. We examine this model in two different cases: single-cell $QB$ and two-cell $QB$. Compared to previous results in Ref. \cite{14} with independent environments, our results demonstrate that the battery has the capability to charge in both non-Markovian (without dipolar interactions) and Markovian (with dipolar interaction and driving fields) dynamics. Also, we find that it is essential to work in an underdamped regime and/or employing driving fields to store more energy in Markovian reservoirs.

Moreover, we figure out a way to achieve a stable N-cell $QB$ by taking into account subradiant states, laser fields and an intermediate regime. However, this is not a new issue and so far many protocols have been developed to stabilize the charging status of $QB$s 
 such as the benefit of memory effects \cite{14}, using dark states \cite{15},  adiabatic protocols \cite{17,18},  Zeno effect \cite{sq}, etc. Finally, we suggest an optical experimental setup to measure ergotropy in the first scenario. 

The rest of the paper is organized as follows. We investigate the open $QB$ in a common environment under two different scenarios. We present the first model: non-Markovian dynamics in Sec.~\ref{II} and the second model: Markovian dynamics in Sec.~\ref{III}. The optical setup is discussed in  Sec.~\ref{IV}. The conclusion is summarized in Sec.~\ref{V}.

 \begin{figure}
\includegraphics[scale=0.5]{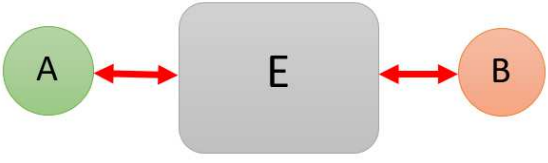}
\caption{Schematic representation of the model considered in Sec. II.}
\label{fig1}
\end{figure}

\section{First scenario: non-Markovian dynamics}\label{II}
We consider a model in this work that consists of three subsystems: a quantum charger A,  a quantum battery B, and a common environment that acts as a mediator between A and B (shown in Fig.~\ref{fig1}). We describe the quantum battery-charger model as a two-qubit system, where the qubits are subjected to the  common zero temperature bosonic reservoir in a vacuum state. Each battery cell is modeled as a two-level system with excitation state $\vert e\rangle$ and ground state $\vert g\rangle$, where we assume two qubits have the same transition frequency $\omega_{\text{A}}=\omega_{\text{B}}=\omega_{0}$. The total Hamiltonian of the system in the rotating-wave approximation (RWA) described as follows (with $\hbar=1$) \cite{20,21} 
\begin{align}\label{e1}
H=H_{0}+ f(t)~H_{I},
\end{align}
in which the total free Hamiltonian is given by
\begin{align}\label{e20}
H_{0} =H_{A}+H_{B}+H_{E}=
\sum_{j=\lbrace \text{A},\text{B}\rbrace}\omega_{0}\sigma^{+}_{j}\sigma^{-}_{j}+\sum_{k}\omega_{k}a^{\dagger}_{k}a_{k},
\end{align}
where the first term denotes the free Hamiltonian of the two-qubit system, with $\sigma^{+}_{j}$ and $\sigma^{-}_{j}$ being the Pauli raising and lowering operators for $j$-th qubits, respectively. The second term in the above equation is the free Hamiltonian of the reservoir, with $a_{k}$ and $a^{\dagger}_{k}$ being the annihilation and creation operators of the $k$-th mode of the field with frequency $\omega_{k}$. The last term in the Eq.~(\ref{e1}) describes the interaction of the system with the reservoir
\begin{align}
&H_{I}=H_{AE}+H_{BE}=\sum_{k}g_{k}\mu_{1}\left(\sigma^{+}_{\text{A}}a_{k}+\sigma^{-}_{\text{A}}a^{\dagger}_{k} \right)\nonumber\\
&+\sum_{k}g_{k}\mu_{2}\left(\sigma^{+}_{\text{B}}a_{k}+\sigma^{-}_{\text{B}}a^{\dagger}_{k}\right),
\end{align}
where $g_{k}\mu_{i}$, $(i=1,2)$ is the coupling constant between the charger/battery and the $k$-th mode of the field, in which $\mu_{i}$ is a dimensionless real parameter. The relative interaction strength is defined as $c_{i}=\mu_{i}/\mu_{T}$ and the collective coupling constant as $\mu_{T}=(\mu^{2}_{1}+\mu^{2}_{2})^{\frac{1}{2}}$. Here, we assume that $\mu_{1}$ and $\mu_{2}$ can be different, so in our discussion, we can consider the different effective coupling of the reservoir to the battery and the charger. 

\begin{figure*}[t!]
	\centering
	\subfloat{\includegraphics[scale=0.335]{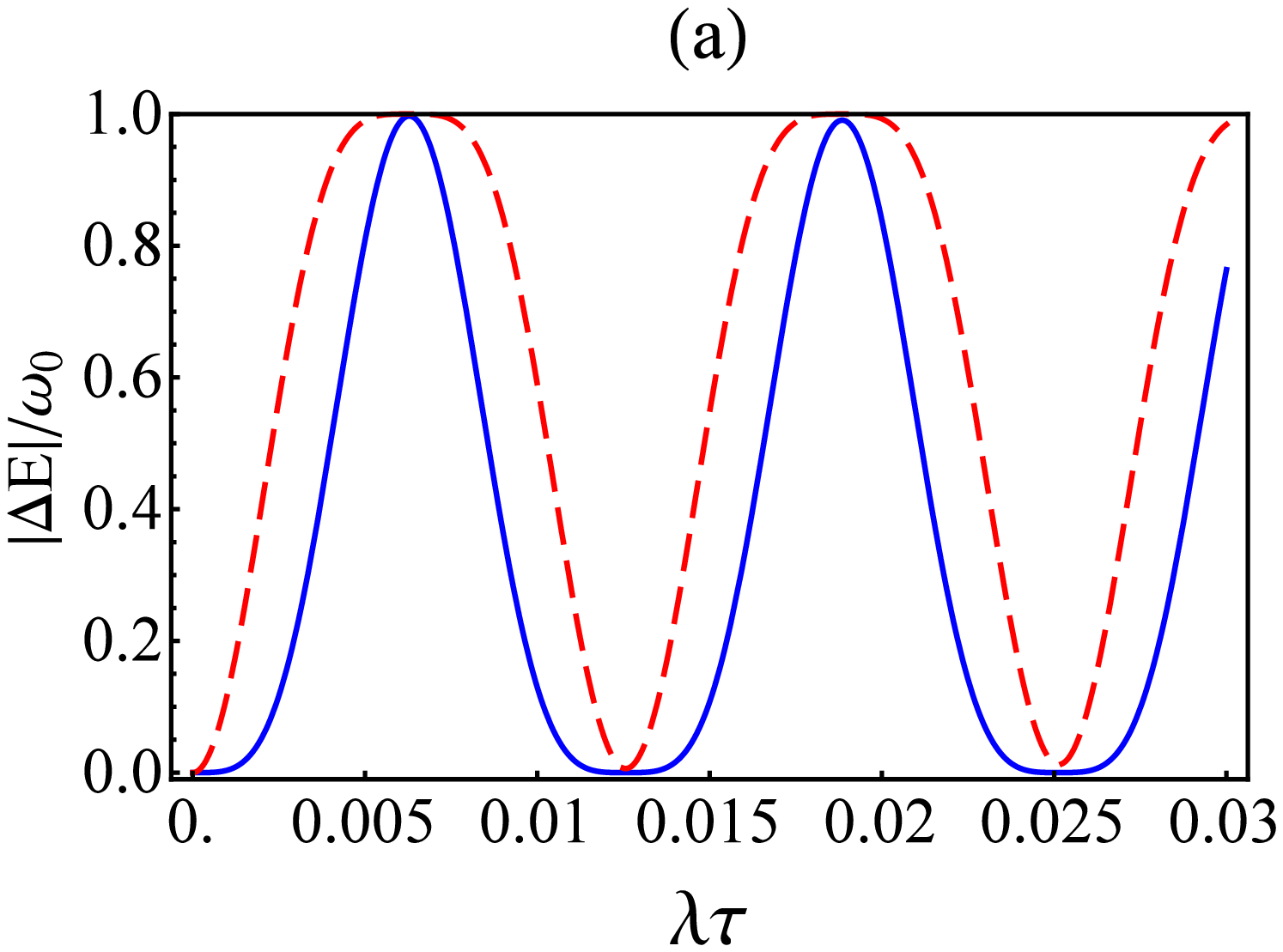}\label{fig2a}}~~~\subfloat{\includegraphics[scale=0.33]{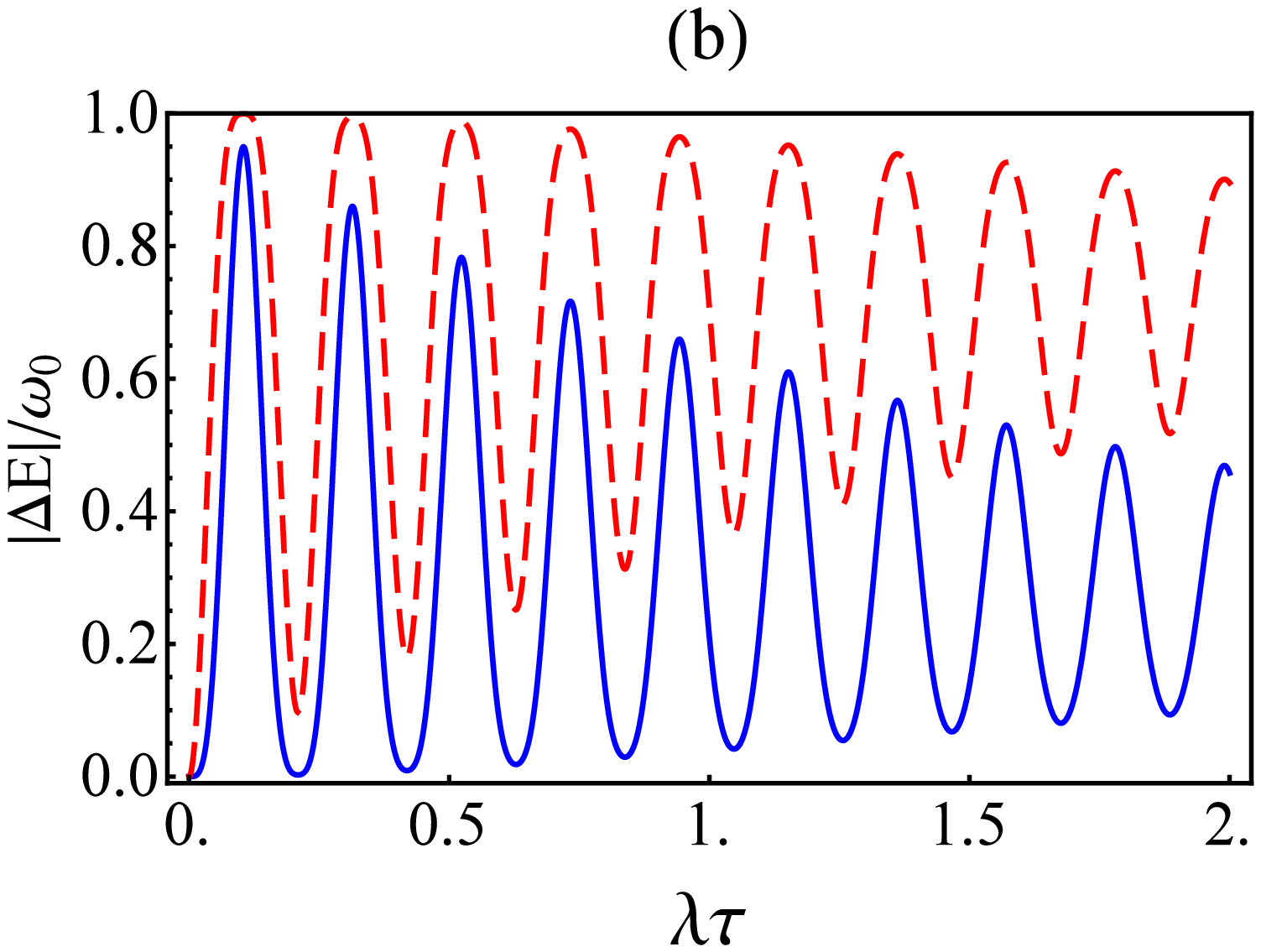}\label{fig2b}}~~~
	\subfloat{\includegraphics[scale=0.33]{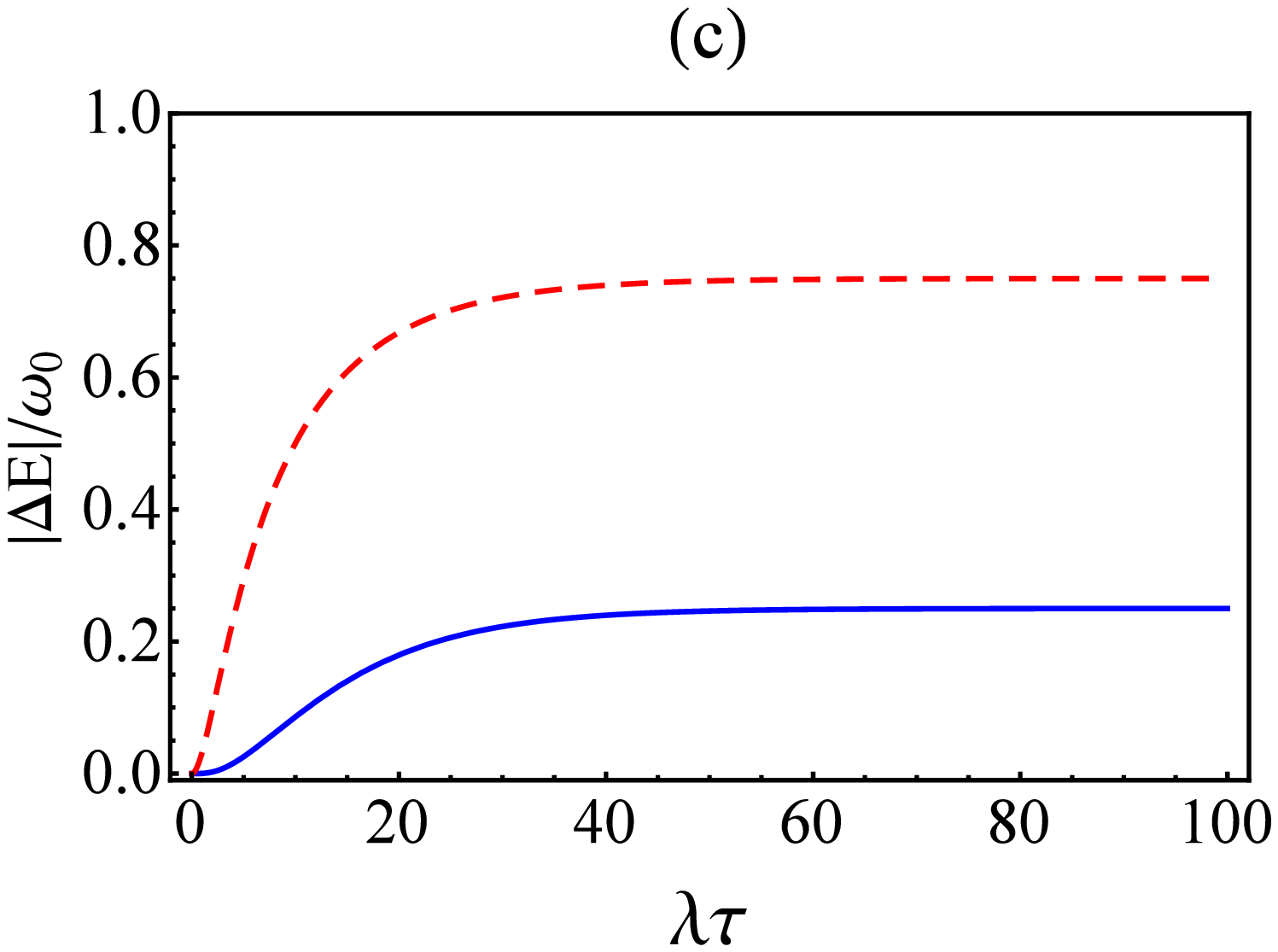}\label{fig2c}}
	\caption{Dynamics of $ \vert \Delta E\vert /\omega_{0}$ as a function of the dimensionless
quantity $\lambda \tau $. Dotted red line depicts $\vert\Delta E_{A}\vert/\omega_{0}$ and solid blue line shows $\Delta E_{B}/\omega_{0}$. The initial state is $\vert\Phi(0)\rangle=\vert e\rangle_{A}\vert g\rangle_{B}\otimes \ket{0}_{\mathcal{E}} $ and $c_{1}=1/\sqrt{2}$. We have: (a) $\mathcal{R}=500$, (b) $\mathcal{R}=30$, and (c) $\mathcal{R}=0.3$.}
	\label{fig2}
\end{figure*}

In Eq.~\ref{e1}, $f(t)$ is a dimensionless function that it equals to $1$ for $t\in[0, \tau[$ and $0$ elsewhere, which it is used to
switch interactions on or off, and $\tau$ presents the charging time of the $QB$.
We assume that for $t < 0$, the charger and the $QB$ are isolated and do not interact with the environment. At time $t = 0$, $\text{A}$ is connected to $\text{E}$ as well as $\text{B}$ to $\text{E}$ by switching on $H_{\text{AE}}$ and $H_{\text{BE}}$, respectively. 
In this scenario, $\text{A}$ and $\text{B}$ do not interact with each other. Since $[H_{\text{B}}+H_{\text{E}}, H_{\text{BE}}] \neq 0$ and  the environment is in its ground state at $t=0$, accordingly, maybe the final energy of the $QB$ will not only originate from the charger $\text{A}$ but also from the thermodynamic work of turning the interactions “on/off” at switching times \cite{13,26}. Hence,  a portion of the charger energy moves to $QB$ with the mediation of the environment in the time window $[0, \tau[$. Ultimately, at time $\tau$  the charger and the $QB$ are again isolated by disconnecting $\text{A}$ from $\text{E}$ as well as $\text{B}$ from $\text{E}$. Since infinitely many degrees of freedom of the environment, here, we do not discuss the amount of the thermodynamic work cost but we will show the relation between the charger energy and the $QB$ energy for different couplings in Figure.~\ref{fig2}.

We emphasize that in our model there is no direct coupling between the charger and $QB$ in Eq.~(\ref{e1}), accordingly, we study wireless charging of the $QB$ where the environment plays a mediated role in the charging process.

\subsection{Wireless charging process}
Here, we are interested in situations where energy can be transferred from the charger to the $QB$ or in a similar way from the $QB$ to a consumption center. We intend to calculate the one excitation time
evolution of the total system when the reservoir is initially in a vacuum state. We choose the initial state of the whole system as follows
\begin{equation}\label{e2}
\vert\Phi(0)\rangle=[\nu_{01}\vert e\rangle_{\text{A}}\vert g\rangle_{\text{B}}+\nu_{02}\vert g\rangle_{\text{A}}\vert e\rangle_{\text{B}}]\otimes\vert 0\rangle_{\mathcal{E}},
\end{equation}
where $\vert 0\rangle_{\mathcal{E}}$ is the vacuum state of the reservoir and $\nu_{0i}$, $(i=1,2)$ are the probability amplitudes. Now, in the lack of Born-Markov approximation \cite{25}, one can exactly obtain the evolved state in the basis spanned by single excitation states as 
\begin{align} \label{e3}
\vert\Phi(t)\rangle &=[\nu_{1}(t)\vert e\rangle_{\text{A}}\vert g\rangle_{\text{B}}+\nu_{2}(t)\vert g\rangle_{\text{A}}\vert e\rangle_{\text{B}}]\otimes\vert 0\rangle_{\mathcal{E}}\nonumber\\
& + \sum_{k}\nu_{k}(t)\vert g\rangle_{\text{A}}\vert g\rangle_{\text{B}}]\otimes\vert 1_{k}\rangle_{\mathcal{E}}.
\end{align}
By solving integrodifferential equations for the above probability amplitudes and using the continuum limit for the spectrum of the environment as well as Laplace transform and its inverse, the probability amplitudes become \cite{20,21}
\begin{align} \label{e4}
&\nu_{1}(t) = \left[c^{2}_{2}+c^{2}_{1}~\mathcal{\kappa}(\tau)\right]\nu_{01} -c_{1}c_{2}\left[1-\mathcal{\kappa}(t)\right]\nu_{02}, \nonumber\\
&\nu_{2}(t) =-c_{1}c_{2}\left[1-\mathcal{\kappa}(t)\right] \nu_{01}+ \left[c^{2}_{1}+c^{2}_{2}~\mathcal{\kappa}(t)\right]\nu_{02}.
\end{align}
Consider the environment as an electromagnetic field inside a cavity with non-ideal mirrors that the spectral density of the cavity field takes the Lorentzian form as follows \cite{J}
\begin{equation} 
J(\omega)=\frac{\xi^{2}\lambda}{\pi[(\omega-\omega_{0})^{2}-\lambda^{2}]}\text{ , }
\end{equation}
in which $\lambda$ being the width of the spectrum where $\lambda^{-1}$ is the correlation
time of the environment and $\xi$ is the effective coupling strength related to the vacuum Rabi frequency $R=\xi~\mu_{T}$. Therefore, one can define the dimensionless parameter $\mathcal{R}=R/ \lambda$ to distinguish the strong coupling regime ($\mathcal{R}\gg 1$) from the weak one  ($\mathcal{R}\ll 1$). With regard to the above Lorentzian spectral density, $\mathcal{\kappa}(\tau)$ is characterized as
\begin{equation}\label{e5}
\mathcal{\kappa}(t)=e^{-\lambda t/2}\left( \cosh (\frac{\chi~t}{2})+\frac{\lambda}{\chi}\sinh (\frac{\chi~ t}{2})\right) ,
\end{equation}
in which $\chi=\sqrt{\lambda^{2}-4 R^{2}}$\cite{20,21}.

 \begin{figure*}[t!]
	\centering
	\subfloat{\includegraphics[scale=0.4]{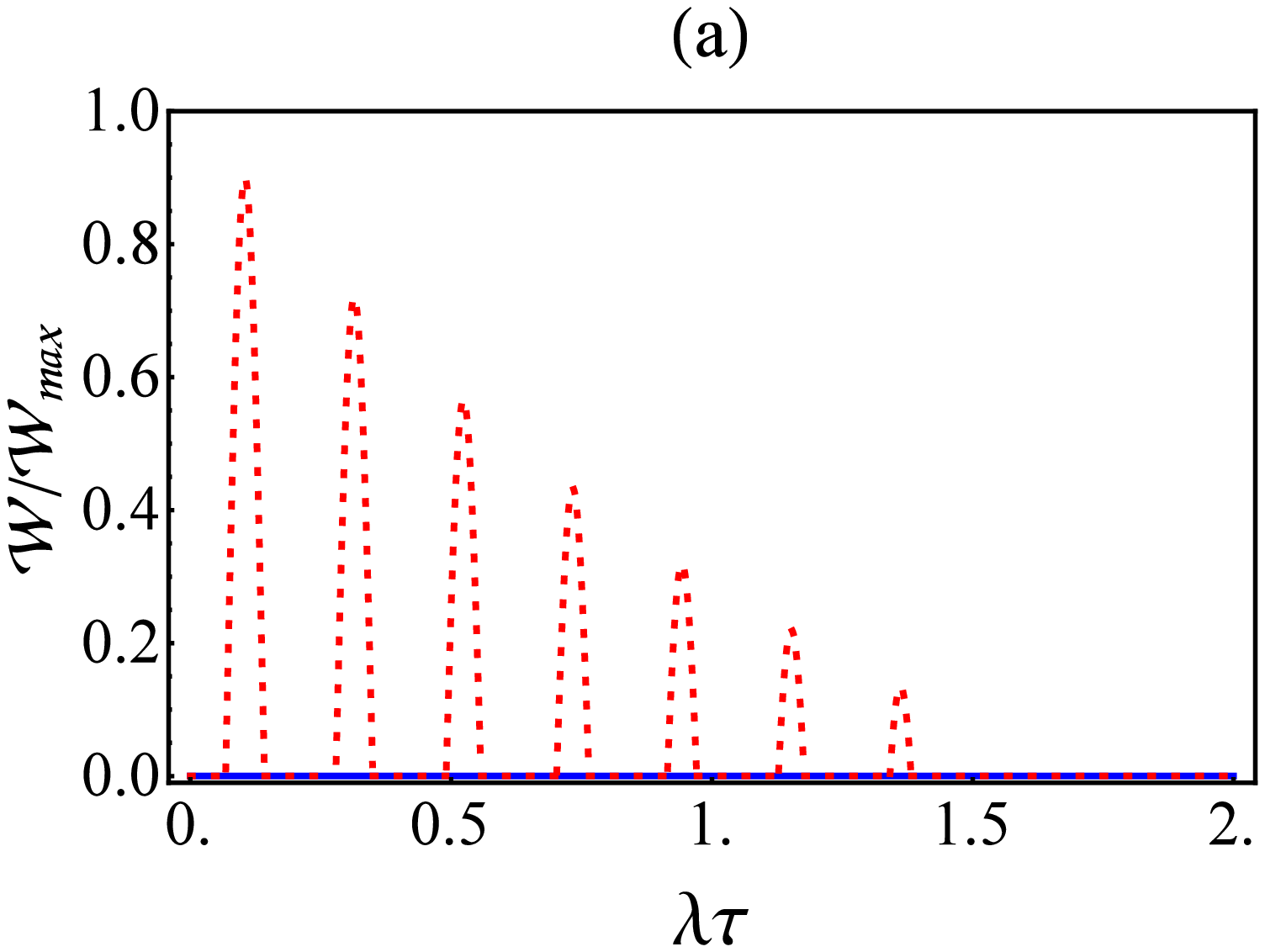}\label{fig3a}}~~~~~~~\subfloat{\includegraphics[scale=0.4]{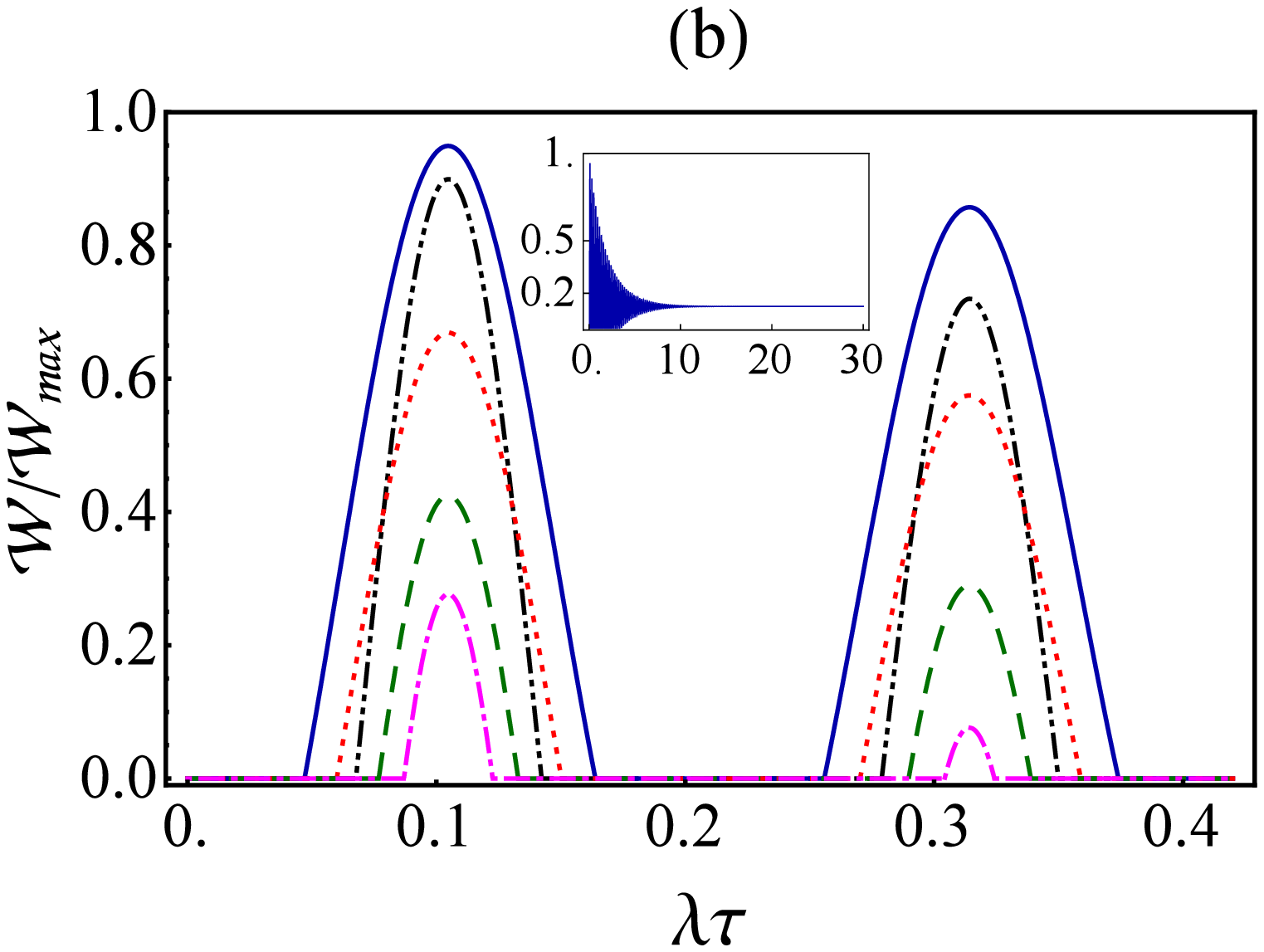}\label{fig3b}}~
	\caption{Dynamics of $\mathcal{W}/\mathcal{W}_{max}$ as a function of the dimensionless
quantity $\lambda \tau$. (a) Dotted red line depicts $\mathcal{R}=30$ and solid blue line $\mathcal{R}=0.3$. The initial state is $\ket{\Phi(0)}=\vert e\rangle_{A}\vert g\rangle_{B}\otimes\ket{0}_{\mathcal{E}}$. (b) Here the entangled initial state of the qubits is assumed as Eq.~(\ref{e9}) and  $\mathcal{R}=30$ is considered. Solid darker blue line shows $(c_{1} = \alpha_{-}=\sqrt{3}/2)$, dot-dot-dashed black line $(c_{1} = \alpha_{-}=1/\sqrt{2})$, dotted red line $(c_{1} =1/\sqrt{2},~ \alpha_{-}=0.92)$, dashed green line $(c_{1} =\sqrt{3}/2, ~\alpha_{-}=0.5)$, dot-dashed  magenta line $(c_{1} =1/\sqrt{2}, ~\alpha_{-}=0.2)$.}
	\label{fig3}
\end{figure*}
Note that the Markovian dynamics occurs when the environmental correlation time is very less than the relaxation time of the qubits. However, strong coupling interactions along with low-temperature reservoirs can cause non-Markovian evolution as well as the emergence of memory effects and revival of quantum properties \cite{29}. In our model, non-Markovian dynamics arises for $\mathcal{R}\gg 1$.

In the following, according to Eq.~\ref{e3}, the reduced density operator of the $QB$ and the charger at $t=\tau$ can be written as 
\begin{subequations}
	\begin{align}
\rho_{\text{B}}(\tau) &= \vert\nu_{2}(\tau)\vert^{2} \ket{e}\bra{e}_{\text{B}} +  \left[1-\vert\nu_{2}(\tau)\vert^{2}\right] \ket{g}\bra{g}_{\text{B}}  \text{ ,} \label{QBdensityMatrix} \\
	\rho_{\text{A}}(\tau) &= \vert\nu_{1}(\tau)\vert^{2} \ket{e}\bra{e}_{\text{A}} +  \left[1-\vert\nu_{1}(\tau)\vert^{2}\right] \ket{g}\bra{g}_{\text{A}}  \text{.}
	\end{align}
\end{subequations}
It is straightforward to obtain the internal energy of the charger/battery, i.e., $E_{\text{A/B}}(\tau)\!=\!tr[\rho(\tau)_{\text{A/B}}H_{\text{A/B}}]$, using the above density matrixs. So, we have 
\begin{equation}\label{e21}
E_{\text{A}}(\tau)=\omega_{0}\vert \nu_{1}(\tau) \vert^{2}\text{ , } ~~E_{\text{B}}(\tau)=\omega_{0}\vert \nu_{2}(\tau) \vert^{2}.
\end{equation}
To study the relationship between the charger and the battery energy, we investigate the internal energy changes. Because of
in our example we suppose that the battery is initially empty and the charger has the maximum energy, i.e., $\vert\Phi(0)\rangle=\vert e\rangle_{A}\vert g\rangle_{B}\otimes \ket{0}_{\mathcal{E}}$ (where $\nu_{01}=1$ and $\nu_{02}=0$),  we then define the
amount of energy that the charger loses at the end of charging process as $\vert\Delta E_{\text{A}}(\tau)\vert=\vert E_{\text{A}}(\tau)-E_{\text{A}}(0)\vert$ and the amount of energy that the battery obtains as $\Delta E_{\text{B}}(\tau)=E_{\text{B}}(\tau)-E_{\text{B}}(0)$. 
Notice that $\Delta E_{\text{A}}(\tau)$ is always negative and we have plotted the absolute value here.

In Fig.~\ref{fig2}, we have plotted the dynamics of $\vert\Delta E_{\text{A/B}}(\tau)\vert$ in respect of the dimensionless time $\lambda\tau$  for $c_{1} =1/\sqrt{2}$. At $t=0$, the charger energy is maximum $E_{\text{A}}(0)=\omega_{0}$ and the battery is empty $E_{\text{B}}(0)=0$. It is considered $\mathcal{R}=500$ in Fig.~\ref{fig2a}, $\mathcal{R}=30$ in Fig.~\ref{fig2b}, and $\mathcal{R}=0.3$ in Fig. \ref{fig2c}.  

As can be seen in the presence of strong coupling regime (non-Markovian dynamics) there are some oscillations in $\vert \Delta E_{\text{A/B}}(\tau)\vert$ in Figs.~\ref{fig2a} and \ref{fig2b} while this is not the case in Fig.~\ref{fig2c}. Fig.~\ref{fig2a} implies that the total energy of the charger can transfer to the battery via the environment in a very strong coupling regime, where we have $\vert\Delta E_{\text{A}}\vert= \Delta E_{\text{B}}$  at the peaks in the plot. But we observe from Figs.~\ref{fig2b} and \ref{fig2c} that the energy in the battery is fewer than energy that is released from the charger, $\Delta E_{\text{B}}(\tau)< \vert\Delta E_{\text{A}}(\tau)\vert$. Hence, one can conclude that the rest of the charger energy remains among  degrees of freedom of the environment  as well as correlations or maybe it is transferred to the outside of the total system through the sudden quench of the interaction Hamiltonians.

To characterize the maximal amount of energy that can be extracted from a $QB$ at the end of the charging process under the cyclic unitary operations, the ergotropy is introduced as \cite{e1,e2} 
\begin{eqnarray}\label{e19}
\mathcal{W}(\tau)=Tr (\rho_{B}(\tau)~H_{B})-  Tr(\sigma_{\rho_{B}}~H_{B}).
\end{eqnarray}
in which $H_{B}$ and $\rho_{B}$ are the Hamiltonian and the state of the battery, respectively. $ \sigma_{\rho_{B}}$ is called the passive state of $\rho_{B}$ with the zero extractable energy by cyclic unitary operations \cite{e1}.

Then, according to Eqs.~(\ref{QBdensityMatrix}) and (\ref{e19}) the ergotropy can be obtained as
\begin{eqnarray}\label{e7}
\mathcal{W}(\tau)=\omega_{0}(2\vert \nu_{2}(\tau)\vert^{2}-1) \Theta(\vert \nu_{2}(\tau)\vert^{2}-\frac{1}{2}),
\end{eqnarray}
in which $ \Theta (x-x_{0}) $ is the Heaviside function. Where we have $\mathcal{W}_{max}=\omega_{0}$.

At this point, let us consider an initial state as
\begin{equation}\label{e9}
\vert\Phi(0)\rangle=\left( \alpha_{-}\vert\varphi_{-} \rangle + \alpha_{+}\vert\varphi_{+}\rangle\right) \otimes \ket{0}_{\mathcal{E}} ,
\end{equation}
where $\alpha_{\pm}=\langle\varphi_{\pm}\vert\varphi(0)\rangle$ and  $ \vert\varphi_{+}\rangle= c_{1}\vert e\rangle_{A}\vert g\rangle_{B}+c_{2}\vert g\rangle_{A}\vert e\rangle_{B} $. Also, $\vert\varphi_{-}\rangle $ is
a subradiant, decoherence-free state of the Hamiltonian (\ref{e1}), that does not decay in time. Such states are obtained in cases where   the two atoms have the same Bohr frequency and takes the following form \cite{20,21}
\begin{equation}\label{e8}
\vert\varphi_{-}\rangle= c_{2}\vert e\rangle_{A}\vert g\rangle_{B}-c_{1}\vert g\rangle_{A}\vert e\rangle_{B},
\end{equation}
with the relative interaction strength $c_{i}~(i=1,2)$. According to Eqs.~(\ref{e4}) and (\ref{e9}), the analytical solution for the amplitudes $\nu_{1}(t)$ and $\nu_{2}(t)$ can be written in the following simple form \cite{20,21}
\begin{eqnarray}\label{e10}
&\nu_{1}(t) &=c_{2}\alpha_{-}+c_{1}\mathcal{\kappa}(t)\alpha_{+}\nonumber\\
&\nu_{2}(t)&=-c_{1}\alpha_{-}+c_{2}\mathcal{\kappa}(t)\alpha_{+}.
\end{eqnarray}
Thus, the amount of the ergotropy depends on the specific initial state ($ c_{j}, \alpha_{\pm}$) and on the value of the coefficient $\mathcal{\kappa}(\tau)$.
 
In the following, we study the dynamical behavior of ergotropy for different initial states in both the weak and the strong coupling regimes as a function of $\lambda \tau$. The time evolution of ergotropy (as a multiple of $\mathcal{W}_{max}$) for initial states  $\ket{\Phi(0)}=\vert e\rangle_{A}\vert g\rangle_{B}\otimes\ket{0}_{\mathcal{E}}$ and Eq.~\ref{e9} (for different coefficients of $c_{1}$ and $\alpha_{-}$) is plotted in  Figs.~\ref{fig3a} and \ref{fig3b}, respectively. Where in the former the total state is separable and the battery is empty and the charger energy is maximal, whereas in the latter there is an amount of entanglement in the battery-charger state.

In Fig.~\ref{fig3a}, dashed red line shows a strong coupling between the battery-charger system and environment by considering $\mathcal{R}=30$. It can be easily seen that for $\lambda \tau \simeq 0.1$ the ergotropy is $\mathcal{W}\approx 0.9\mathcal{W}_{max}$, whereas for $\lambda\tau > 1.5$ it vanishes. By contrast, when $\mathcal{R}=0.3$ which corresponds to the weak coupling regime reported by solid blue line in Fig.~\ref{fig3a}, there is an amplification of the decoherence effects since we have no backflow of information from the environment and ergotropy always zero for all values $\lambda \tau$. Comparing Fig.~\ref{fig3a} with Fig.~\ref{fig2c}, one can see that the battery has some energy while its ergotropy is always zero for the weak interaction. Also, similar behavior can be observed for strong coupling regime that ergotropy is zero at some time intervals while its energy is not zero.

\begin{figure*}[t!]
	\centering
	\subfloat{\includegraphics[scale=0.4]{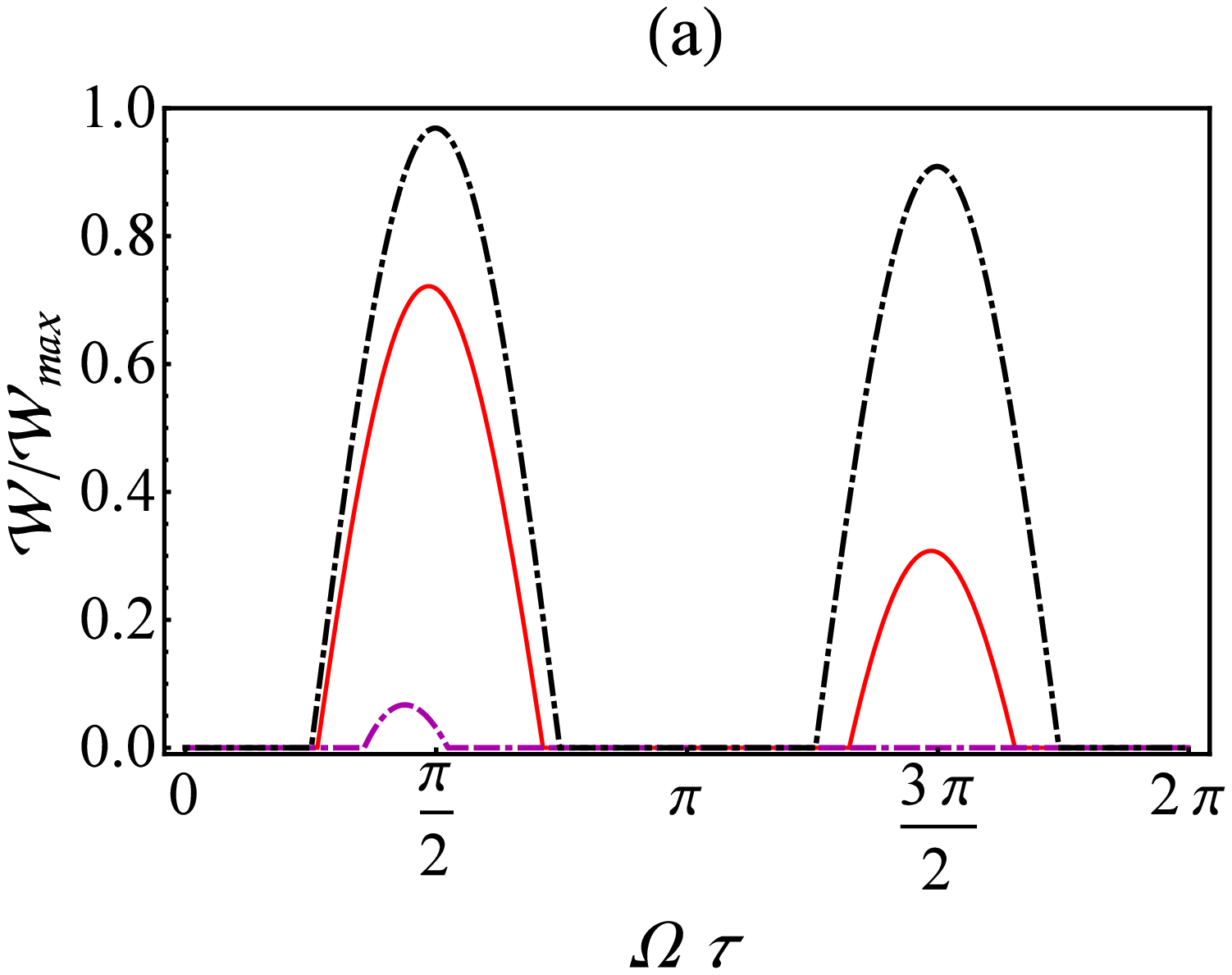}\label{fig4a}}~~~~~~~\subfloat{\includegraphics[scale=0.4]{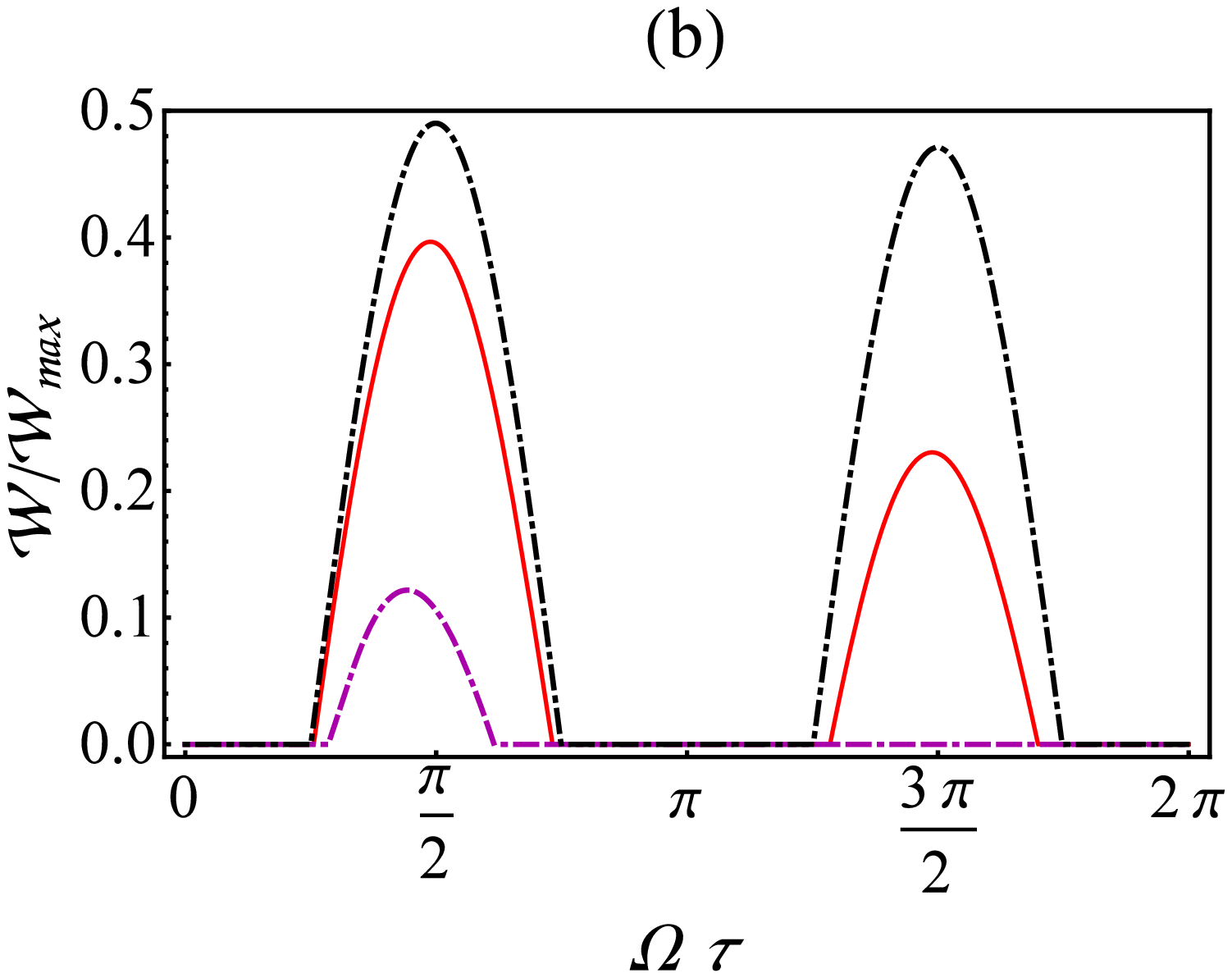}\label{fig4b}}~
	\caption{ Single-cell QB. Time evolution of $\mathcal{W}/\mathcal{W}_{max}$  as a function of the dimensionless
quantity $\Omega \tau$ for $\gamma=0.9 \Gamma$, and $ l_{1}= l_{2}=0$. Dot-dashed-dashed purple line exhibits $(\Gamma=0.5\Omega) $, solid red line $( \Gamma=0.1 \Omega) $, dot-dashed black line $( \Gamma=0.01 \Omega)$. (a) The initial state is assumed $\vert\varphi\rangle=\vert e\rangle_{\text{A}}\vert g\rangle_{\text{B}}$. (b) The initial entangled state is chosen with $\alpha_{-}=c_{1}=\sqrt{3}/2$.}
	\label{fig4}
\end{figure*}

In Fig.~\ref{fig3b}, the ergotropy dynamics is depicted for $\mathcal{R}=30$ and several pairs of parameters $(c_{1},\alpha_{-})$. Here, the solid blue line, dot-dot-dashed black line, dotted red line, dashed green line and dot-dashed magenta line represents $(\sqrt{3}/2,~\sqrt{3}/2)$, $(1/\sqrt{2},~1/\sqrt{2})$, $(1/\sqrt{2},~0.92)$, $(\sqrt{3}/2,~0.5)$, and $(1/\sqrt{2},~0.2)$, respectively. 
One can clearly see that the strong coupling regime makes a non-negligible contribution to the dynamics of charge for $\lambda \tau\simeq 0.1$, where the battery is almost fully charged  with $\mathcal{W}\approx 0.95\mathcal{W}_{max}$. The ergotropy dynamics in long time limit is shown via the inset of Fig.~\ref{fig3b}, one can see that it tends to $0.125\mathcal{W}_{max}$ for sufficiently large times. This happens when there is an amount of entanglement in the initial state, while the ergotropy is zero for long times in Fig.~\ref{fig3a}. 
Finally, we have investigated the time evolution of the ergotropy for the case $\mathcal{R}=0.3$ (its plot has not been shown here), in which the ergotropy reaches approximately to the amount of $0.125 \mathcal{W}_{max}$ for $c_{1} =\alpha_{-}=\frac{\sqrt{3}}{2}$, however, it is zero for other initial states in Fig.~\ref{fig3b}. Moreover, the ergotropy is always zero in the weak coupling regime with $\mathcal{R}=0.1$.

As a result, in order to extract a desirable work form the battery it is essential to have a strong interaction between the battery-charger system and the environment against the case of  individual environments in Ref. \cite{14}. This is an important advantage to realize more operational batteries compared to others \cite{13,Ba,14}.

\section{ Second scenario: Markovian dynamics}\label{III}
In this section, we discuss the situation under which the Born-Markov and rotating-wave approximations are employed, therefore, the master equation of the two-qubit system by applying the external laser fields can be expressed as (with $\hbar=1$)\cite{22}
\begin{align}\label{e12} 
&\frac{\partial\rho(t)}{\partial t}=-i[H_{s},\rho(t)] -i f(t)[H_{d}+H_{L},\rho(t)]\\ \nonumber
&- f(t)\frac{1}{2}\sum_{i,j=1}^{2}\Gamma_{ij}(\rho(t)\sigma^{+}_{i}\sigma^{-}_{j}+
\sigma^{+}_{i}\sigma^{-}_{j}\rho(t)-2\sigma^{-}_{j}\rho(t)\sigma^{+}_{i}),
\end{align}
in which
\begin{equation}\label{e13}
H_{s}=\sum_{i=1}^{2}(\omega_{0})\sigma^{+}_{i}\sigma^{-}_{i},~~~H_{d}=\sum^{2}_{i\neq j , j=1}\Omega_{ij}\sigma^{+}_{i}\sigma^{-}_{j},
\end{equation}
the first term describes the free Hamiltonian of the two-qubit system where the second term represents the environment-induced coherent (dipole-dipole) interaction between the qubits with coupling $\Omega_{ij}~(i\neq j)$ and $H_{L}$ denotes the Hamiltonian of the external fields with frequency $\omega_{L}$ and the Rabi frequency $l_{i}$ (i=1,2) that given by
\begin{equation}\label{e14}
H_{L}=-\frac{1}{2}\sum_{i=1}^{2}l_{i}[\sigma^{+}_{i}e^{i(\omega_{L}t)}+\sigma^{-}_{i}e^{-i(\omega_{L}t)}].
\end{equation} 
In Eq.~(\ref{e12}), the parameters $\Gamma_{ij}$  are spontaneous emission rates where $\Gamma_{i} =\Gamma_{ii}$ is the individual spontaneous emission rate of the $i$-th qubit, and $\Gamma_{ij}=\Gamma_{ji}~(i \neq j)$ is collective spontaneous emission rate due to the coupling between the qubits through the environment. Notice that the collective interactions between the qubits leads to the modified dissipative decay rates and the coherent coupling $\Omega_{ij}$ \cite{22}. It has been demonstrated that both the collective parameters $\Gamma_{ij}$ and $\Omega_{ij}$ are dependent on the interatomic separation. As an example, for large separations i.e., $ r_{12}\gg\overline{\lambda}$ (with the resonant wavelength $\overline{\lambda}$), we have $\Gamma_{ij}=\Omega_{ij}\approx 0$  \cite{22}.
To be specific, now we suppose $\Omega_{12}=\Omega_{21}=\Omega$,  $\Gamma_{i}=\Gamma$, and $\Gamma_{ij}=\gamma$ for $i\neq j$, respectively. Also, f(t) is a function with similar behavior in Eq.~(\ref{e1}).

\begin{figure*}[t!]
	\centering
	\subfloat{\includegraphics[scale=0.45]{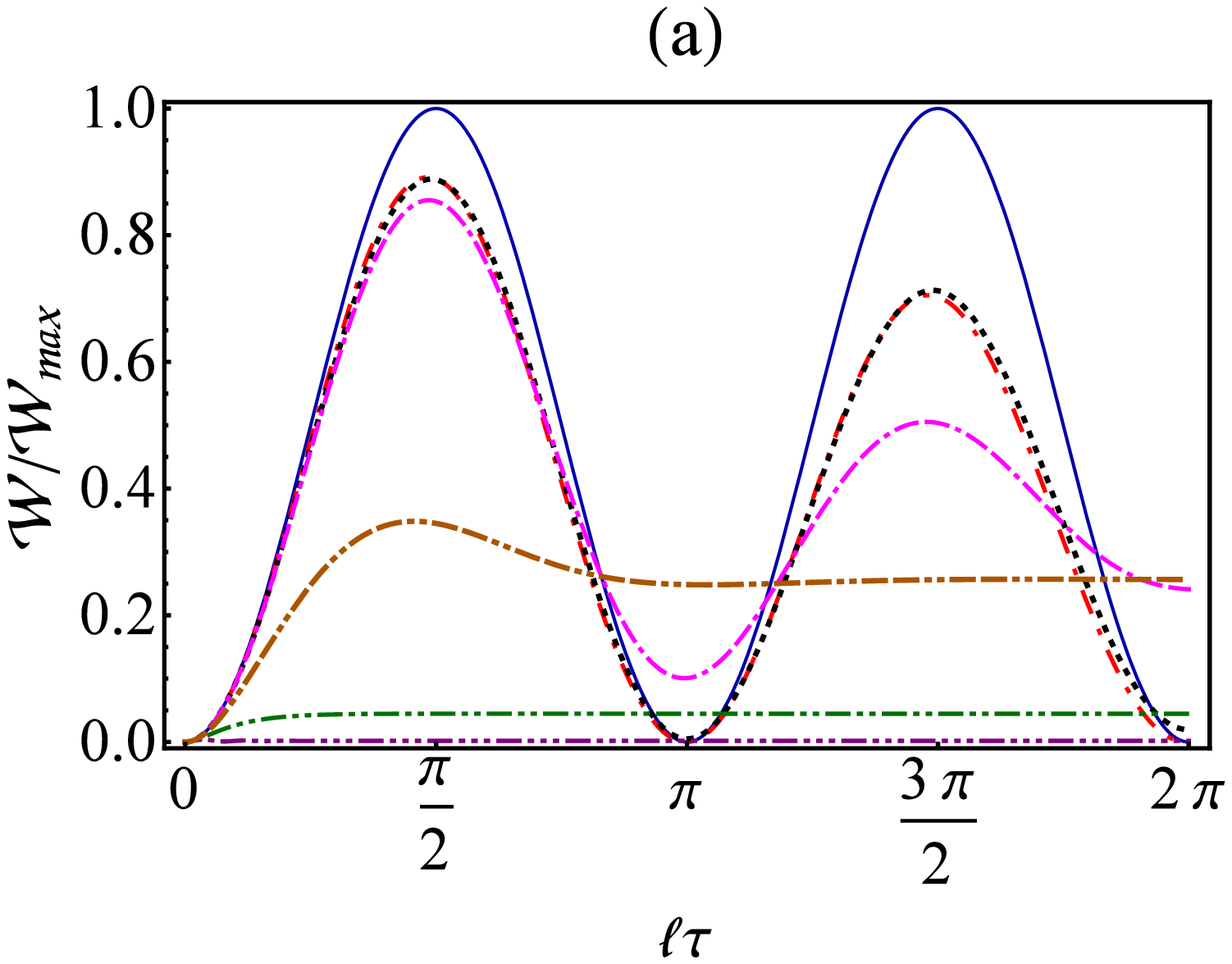}\label{fig5a}}~~~~~\subfloat{\includegraphics[scale=0.48]{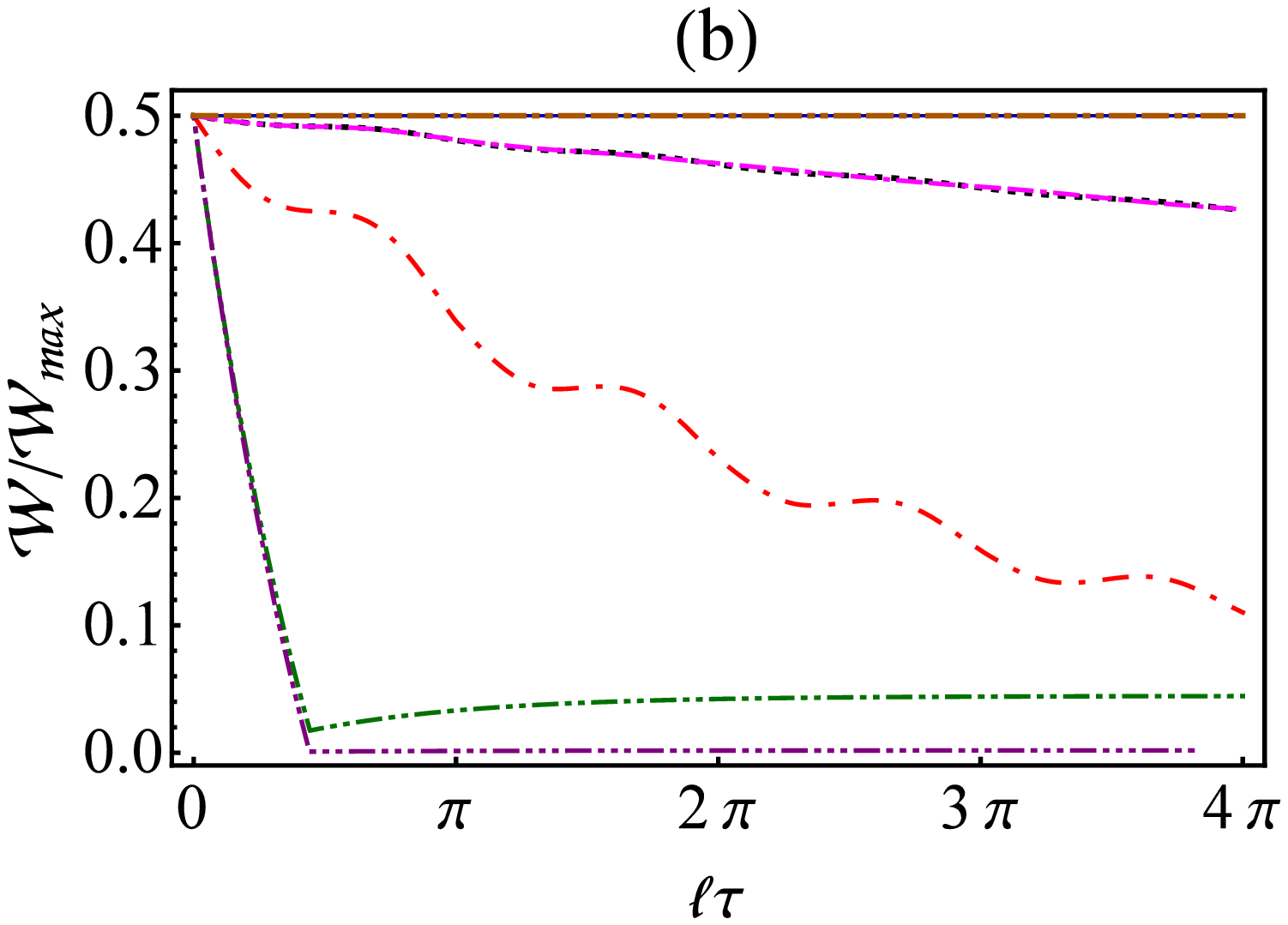}\label{fig5b}}~
	\caption{Two-cell QB. Dynamics of $\mathcal{W}/\mathcal{W}_{max}$ as a function of the dimensionless quantity $l\tau$ by considering $l_{1}= l_{2}=l$, and $\omega_{0}=\omega_{L}=\omega$. Solid darker blue line shows $(\Gamma=\gamma=\Omega=0)$, dot-dot-dashed-dashed darker orange line $(\Gamma=\gamma=\Omega=l)$, dot-dashed red line $(\Gamma=0.1 l,~\gamma=\Omega=0)$, dotted black line $(\Gamma=0.1l, ~\Omega=0.1\Gamma,~ \gamma=0.9\Gamma)$, dot-dashed-dashed magenta line $(\Gamma=0.1l,~ \Omega=5\Gamma,~ \gamma=0.9\Gamma)$, dot-dot-dashed green line $(\Gamma=5 l,~ \Omega=0.1\Gamma,~ \gamma=0.9\Gamma)$, dot-dot-dot-dashed purple line $(\Gamma=5l,~ \Omega=5\Gamma,~\gamma=0.9\Gamma)$. The battery initial state is $\vert\varphi\rangle_{\text{B}}=\vert gg\rangle$ in (a) and $\vert \varphi\rangle_{B}=  \vert\varphi_{-} \rangle$ with $c_{1}=1/\sqrt{2}$ in (b).}
	\label{fig5}
\end{figure*}

To obtain the dynamics of state $\rho$ at a generic time instant $t$, we solve  Eq.~(\ref{e12}) numerically by writing the $\rho$ in the following matrix form
\begin{equation}\label{e15}
\rho(t)=\begin{pmatrix}
   \rho_{11}(t) &  \rho_{12}(t) & \rho_{13}(t)  & \rho_{14}(t)  \\
     \rho_{21}(t) & \rho_{22}(t) &  \rho_{23}(t)  & \rho_{24}(t) \\
      \rho_{31}(t) &  \rho_{32}(t) &  \rho_{33}(t)  & \rho_{34}(t) \\
     \rho_{41}(t) &  \rho_{42}(t) &  \rho_{43}(t)  & \rho_{44}(t) \\
\end{pmatrix},
\end{equation}
in which we have used a basis set as $\lbrace\ket{m}\rbrace_{m\in\lbrace1,...,4\rbrace}$ where $\ket{1}=\ket{1}_{A}\ket{1}_{B}$, $\ket{2}=\ket{1}_{A}\ket{0}_{B}$, $\ket{3}=\ket{0}_{A}\ket{1}_{B}$ and $\ket{4}=\ket{0}_{A}\ket{0}_{B}$.

It is important to notice that compared to the previous section here there is a dipole-dipole interaction between two qubits as well as external driving fields in addition to dissipation effects. Hence, we expect the battery to charge proportionally, despite weak coupling and Markovian evolution.

In the following, one can inquire two models: (i) single-cell $QB$, a charger-battery protocol without any external coherent field as the scenario in the previous section, (ii) two-cell $QB$, we consider two-qubit system as a $QB$ where each of the qubits is charged by a laser.

According to Eqs.~(\ref{e19}) and (\ref{e15}), the analytical expression of the ergotropy for the single-cell model with $H_{B}=\omega_{0}\sigma^{+}_{2}\sigma^{-}_{2}$ can be evaluated as \cite{13}
\begin{eqnarray}\label{e16}
\mathcal{W}(\tau)&=&\frac{\omega_{0}}{2}\lbrace \sqrt{4\vert \rho_{12}(\tau) +\rho_{34}(\tau)\vert^{2}+ (2 [\rho_{11}(\tau)+\rho_{33}(\tau)]-1)^{2}}\nonumber\\
&+&2 [\rho_{11}(\tau)+\rho_{33}(\tau)]-1\rbrace,
\end{eqnarray}
and for the two-cell case with $H_{B}=\sum_{i=1}^{2}\omega_{0}\sigma^{+}_{i}\sigma^{-}_{i}$  as
 \begin{eqnarray}\label{e17}
\mathcal{W}(\tau)&=&\omega_{0}(-2\eta_{1}(\tau)-\eta_{2}(\tau)-\eta_{3}(\tau)\nonumber\\
&+&2 \rho_{11}(\tau)+\rho_{22}(\tau)+\rho_{33}(\tau)),
\end{eqnarray}
in which $\eta_{i}$'s are the eigenvalues of $\rho(\tau)$ such that $\eta_{i}(\tau)\leq\eta_{i+1}(\tau)$.

\subsection{Single-cell QB} 
In this subsection, the first qubit is treated as a charger and the second one as a battery in the absence of external coherent fields, i.e., $l_{1} =l_{2}=0$. Here, the maximum ergotropy is $\mathcal{W}_{max}=\omega_{0}$ as the previous section.

To investigate the influence of the spontaneous emission rates on the ergotropy, we have presented its dynamics with respect to the dimensionless quantity $\Omega \tau$ in Figs. \ref{fig4a} and \ref{fig4b}. Where the decay rate $\gamma$ is assumed to be $0.9 \Gamma$ and dot-dashed-dashed purple line is fixed on $\Gamma=0.5\Omega$, solid red line on $ \Gamma=0.1 \Omega$, and dot-dashed black line on $\Gamma=0.01 \Omega $.
Figure \ref{fig4a} shows the dynamics of the ergotropy for the initial state $\vert\varphi\rangle_{AB}=\vert e\rangle_{\text{A}}\vert g\rangle_{\text{B}}$ of QB-charger system and Fig.~\ref{fig4b} displays it for the initial state $\vert \varphi\rangle_{AB}= \alpha_{-} \vert\varphi_{-} \rangle + \alpha_{+} \vert\varphi_{+} \rangle$ with $\alpha_{-}\!=\! c_{1}\!=\!\sqrt{3}/2$. We observe that the effect of the spontaneous emission on the ergotropy is destructive; the ergotropy decreases with increasing $\Gamma$ with respect to dipole-dipole interaction, and the extractable work may disappear for large spontaneous emission rates. As can be seen, the ergotropy reaches its maximum value at $\Omega \tau=\pi/2$ by reducing the spontaneous emission rate in regard to dipolar interaction (underdamped regime). 
By comparing Fig.~\ref{fig4b} with Fig.~\ref{fig4a} one can see that the initial entanglement between the charger and the battery does not have an effective  effect on the ergotropy behavior in Born-Markov regime as well as in non-Markov case in the previous section. With these considerations and the results of the previous section, it can be suggested that in such models, the initial entanglement is not a useful resource for the optimal charging process of open quantum batteries as it has been demonstrated that entanglement is not an essential resource to optimal work extraction \cite{2, 27}.
  
\subsection{Two-cell QB} 
Now, Let us consider the two qubits as a $QB$, where the lasers are applied with $l_{1}=l_{2}=l$ and $\omega_{0}=\omega_{L}=\omega $. The evolution of the ergotropy as a function of the dimensionless time $l\tau$ is shown in Figs.~\ref{fig5a} and \ref{fig5b} for  different cases which correspond to the regimes obtained by comparing parameters $\Omega$ and $l$ with $\Gamma$. So, one can find four different regimes, where $\Omega\gg \Gamma$ in sense that dipole-dipole interaction is greater than spontaneous emission rate (underdamped regime) and another regime by $l\gg \Gamma$, means that driving external fields are stronger than spontaneous emission rate and vice-versa, i.e., $\Omega\ll \Gamma$ (overdamped regime) and $l \ll \Gamma$. 
The battery initial state is considered as $\vert\varphi\rangle_{\text{B}}=\vert gg\rangle$ (completely empty) and $\vert \varphi\rangle_{B}=  \vert\varphi_{-} \rangle$ (subradiant state) with $c_{1}=1/\sqrt{2}$  in Figs.~\ref{fig5a} and \ref{fig5b}, respectively.  Here, the highest value of the ergotropy is $\mathcal{W}_{max}=2\omega_{0}$, hence we have normalized it to the unit. 

To clarify the above discussion in Fig.~\ref{fig5a}, we first regard a situation where the environmental effects are not present (solid darker blue line), i.e., $ \Gamma=\gamma=\Omega=0 $. It is noteworthy that the $QB$ can be fully charged by lasers and its ergotropy changes periodically over time. 
While is drastically reduced in dot-dot-dashed-dashed darker orange line with $\Gamma=\gamma=\Omega=l$ (an intermediate regime). This can be due to the fact that a large part of the battery's energy destroys by the environment, such that it almost vanishes under conditions presented in dot-dot-dot-dashed purple line and dot-dot-dashed green line. Also, a status with large separations $r_{12}\gg\overline{\lambda}$ is indicated by dot-dashed red line, where one can take $\gamma=\Omega=0$ and $\Gamma=0.1 l$. 

 Moreover, both dotted black line and dot-dashed-dashed magenta line represent $l \gg \Gamma$ regime where the former implies $\Omega\gg \Gamma$ whereas the latter characterizes $\Omega \ll \Gamma$.  We see the ergotropy is nearly $0.9 \mathcal{W}_{max}$ at  $l \tau=\pi /2$ for black line which shows a very similar behavior in accordance with the large separation status. On the other hand, the $ l \ll\Gamma $ regime is illustrated by dot-dot-dashed green line and dot-dot-dot-dashed purple line with $\Omega\gg \Gamma$ and $\Omega \ll \Gamma $, respectively. 
By regarding many different values for $\gamma$, we find that the collective emission decay rate does not play an effective role in ergotropy dynamics. In addition, Fig.~\ref{fig5a} demonstrates that the ratio $\Gamma/l$ is more significant than $\Gamma/\Omega$ because by reducing the former the ergotropy tends to the unit. This implies that the driving external fields play a substantial role in this scenario.

\begin{figure}
\includegraphics[scale=0.5]{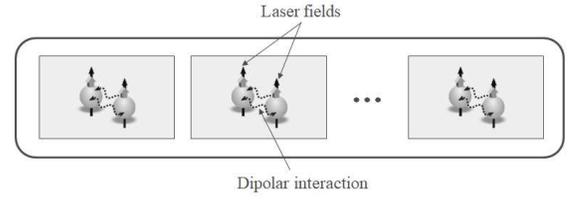} 
\caption{Schematic representation of $N$-cell stable quantum battery by considering $2N$ qubits where every two qubits interact with a common environment and in the presence of driving fields as well as the intermediated regime.}
\label{fig6}
\end{figure} 

As can be seen in Fig.~\ref{fig5b} unlike the previous case, the dot-dot-dashed-dashed darker orange line has coincided with the solid blue line and the ergotropy does not change over time. Indeed, it remains constant for intermediate regime, i.e., $\mathcal{W}= 0.5\mathcal{W}_{max}$ for all times. This indicates the energy of two-cell battery equals to the energy of a single-cell fully charged battery, i.e., $\mathcal{W}=\omega_{0}$. In a more interesting scenario, our results imply that by considering $2N$ qubits such that every two qubits takes into account as a two-cell battery in a common reservoir (see Fig.~\ref{fig6}), then we have $\mathcal{W}= N \omega_{0}$. Notice that this amount of energy is equal to the amount of $N$ single-cell $QB$ that are completely charged.
In context of open quantum batteries the operational quantum batteries must be capable to prevent energy leakage because of decoherence effects of environment  \cite{15,17,18,sq}. Thus, we must study ways to energy trapping and realize stable quantum batteries. It is worth emphasizing that by taking into account $2N$ qubits where every two-qubit is in a subradiant state under the intermediate regime and by applying external fields, we find a stable battery with the amount of energy $N\omega_{0}$ that keeps its energy and is not affected by the destructive effects of the environment. 
Indeed, this strategy for the robust battery can be applied for the first scenario (non-Markovian dynamics) by regarding two qubits as a $QB$ with initial subradiant state $\vert\varphi_{-} \rangle$. We stress that this stable battery may also be examined in a lab, note the considerations in the next section. 
In addition, we observe the ergotropy decays for other regimes. In addition, we discover that the subradiant state $\vert\varphi_{-} \rangle $ is not always a decoherence-free state in Born-Markov approximation.
Also, in $l \gg \Gamma$ regime the battery charge decays to zero monotonically, whiles the battery discharges very fast in the $l \ll\Gamma$ regime. Therefore, the $QB$ is clearly more stable in the presence of a strong external driving field compared to the weak one. Similar to the above, in this case the ratio $\Gamma/l$ is more efficient factor in the stability of quantum batteries. As it increases, the battery discharge time becomes longer.

\section{Experimental study of wireless-like quantum battery} \label{IV}
In this section, we present some general discussions on ergotropy measurement experimentally. To this end, we introduce an optical setup that has been suggested in \cite{23,24}. We define a two-qubit system based on degrees of freedom of a single photon. 
The horizontal and the vertical polarization are regarded as the ground and the excited state of the first qubit, i.e., 
$  \vert H\rangle \equiv  \vert g\rangle $ and $  \vert V\rangle \equiv  \vert e\rangle $.
As well as the first-order Hermitian-Gaussian modes are used for the second qubit, the $HG_{01}$ and the $HG_{10}$ mode are considered as the ground and the excited state of the second qubit, i.e., $  \vert g\rangle \equiv  \vert HG_{01}\rangle \equiv  \vert h\rangle $ and  $  \vert e\rangle \equiv  \vert HG_{10}\rangle \equiv  \vert v\rangle $. 
Here, horizontal and vertical nodal line are shown by $h$ and $v$, respectively. Therefore, the relationship between states of the qubits and degrees of freedoms of the photon is as follows $ \lbrace \vert ee\rangle \equiv  \vert V v\rangle, \vert eg\rangle \equiv  \vert V h\rangle, \vert ge\rangle \equiv  \vert H v\rangle, \vert gg\rangle \equiv  \vert Hh \rangle \rbrace$. In addition, different paths of the photon are considered as
 different states of the environment.

Assume the initial state as $\vert\Phi(0)\rangle=\vert e\rangle_{\text{A}}\vert g\rangle_{\text{B}}\otimes\vert 0\rangle_{\mathcal{E}}=\ket{Vh}_{AB}\ket{0}_{\mathcal{E}}$, then its evolution according to Eq.~(\ref{e3}) can be rewritten as 
$ \vert\Phi(\tau)\rangle=(\nu_{1}(\tau) \ket{Vh}_{AB}+\nu_{2}(\tau)\ket{Hv}_{AB})\ket{0}_{\mathcal{E}}+\sum_{k}\nu_{k}(\tau)\ket{Hh}_{AB}\ket{1_{k}}_{\mathcal{E}}$.

\begin{figure}
\includegraphics[scale=0.5]{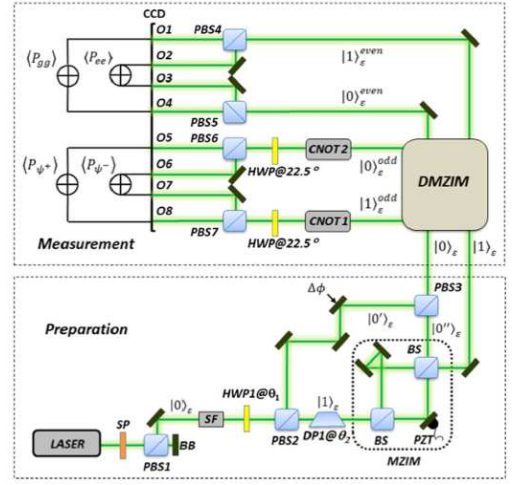} 
\caption{Schematic diagram of optical setup, it is taken from reference \cite{23}. SP (S-wave plate), PBS (polarizing beam splitter), SF (spatial filter), HWP(half wave plate), DP(Dove prism),  MZIM (Mach-Zehnder interferometer with an additional mirror), PZT (piezoelectric
ceramic), DMZIM (double input/output MZIM), BS(beam splitter), CNOT(controlled-NOT gate), CCD camera (charge coupled device camera).}
\label{fig7}
\end{figure}

According to setup shown in Fig.~\ref{fig7}, the entangled state $ 1/\sqrt{2}(\ket{Vh}+\ket{Hv})$ is created when a vertically polarized diode laser beam passes via a SP. Then, PBS reflects beams with vertial polarizaiton while transmits beams with horizontal polarization, accordingly, the state $\ket{Vh}$ ($\ket{Hv}$) is reflected (transmitted) through $PBS1$. By characterizing the path following the SF as the vacuum state of the environment, the initial state is prepared as $\ket{Vh}_{AB}\ket{0}_{\mathcal{E}}$. After applying a half wave plate (HWP1) that it is aligned at an angle $\theta_{1}$ with respect to the vertical polarization, PBS2, Dove prism (DP1@$\theta_{2}$) aligned at an angle $\theta_{2}$ with respect to the vertical orientation, a Mach-Zehnder interferometer with an additional mirror (MZIM) and PBS3, the final state can be written as \cite{23,24}
\begin{eqnarray}\label{e22}
\ket{Vh}_{AB}\ket{0}_{\mathcal{E}}&=& [\cos(2\theta_{1})\ket{Vh}+\sin(2\theta_{2})\sin(2\theta_{1})\ket{Hv}]_{AB}\ket{0}_{\mathcal{E}}\nonumber\\
& +&\sin(2\theta_{1})\cos(2\theta_{2})\ket{Hh}_{AB}\ket{1}_{\mathcal{E}}.
\end{eqnarray}
By comparing the above equation with Eq.~(\ref{e3}), one can simulate $\nu_{1}(\tau)\equiv\cos(2\theta_{1})$, $\nu_{2}(\tau)\equiv\sin(2\theta_{2})\sin(2\theta_{1})$, and $\sum_{k}\nu_{k}(\tau)\equiv \sin(2\theta_{1})\cos(2\theta_{2})$.
Thus, the interaction of the qubits with the environment is accomplished by adjusting two angels $\theta_{1}$ and $\theta_{2}$. 

After the final state is prepared in the preparation circuit, the states in path $\ket{0}_{\mathcal{E}}$ and $ \ket{1}_{\mathcal{E}}$ are steered to the measurement circuit. Then, with modulation the phase difference between arms in a DMZIM,
the paths $\ket{0}_{\mathcal{E}}$ and $ \ket{1}_{\mathcal{E}}$ will be
divided into the even and odd ingredients. In the following, the components  $\ket{1}^{even}_{\mathcal{E}}$ and $\ket{0}^{even}_{\mathcal{E}}$ are directed to PBS4 and PBS5, respectively. The components $ \ket{1}^{odd}_{\mathcal{E}} $ and $ \ket{0}^{odd}_{\mathcal{E}} $ are directed to controlled-NOT gate then a half-wave plate (HWP@$ 22.5^{\circ}$), and finally PBS6 and PBS7, respectively. 
Outputs $O_{1}$ plus $O_{4}$ ($O_{2} $ plus $O_{3}$)  measure the intensity corresponding to state $\ket{Hh}$ ($\ket{Vv}$). Moreover, $O_{5}$ plus $O_{8}$ ($O_{6}$ plus $O_{7}$) measure the intensity corresponding to state $\ket{\psi^{+}} $ ($ \ket{\psi^{-}}$). In the following, by applying a CCD camera the image of all outputs are recorded in a frame. 
$ I_{i} $ is defined as the intensity of the corresponding output $O_{i} $, $ (i=1,2,...,8)$, and the total intensity as $I_{T}=\sum^{8}_{i=1} I_{i}$.
The population of each state $\ket{\phi}$ is given by $\langle P_{\phi}\rangle$ where $\ket{\phi}= \lbrace \ket{ee}, \ket{gg}, \ket{\psi}^{+}, \ket{\psi}^{-}\rbrace$. The populations are specified in respect of intensities as  $\langle P_{\psi}^{+}\rangle= (I_{5}+I_{8})/ I_{T}$, $\langle P_{\psi}^{-}\rangle= (I_{6}+I_{7})/ I_{T} $, $\langle P_{gg}\rangle= (I_{1}+I_{4})/ I_{T}$, and $\langle P_{ee}\rangle= (I_{2}+I_{3})/ I_{T} $ \cite{24}.

Let us consider the case in Fig.~\ref{fig2a} that the battery is fully charged at $ \lambda\tau\approx 0.006 $. It  corresponds to angles $\theta_{1}=\theta_{2}=\pi/4$. Then, in this case we have $\langle P_{ge}\rangle=\langle P_{\psi}^{+}\rangle+\langle P_{\psi}^{-}\rangle $. Hence, the ergotropy according to Eq.~(\ref{e7}) can be calculated experimentally.

Therefore, by measuring the intensity of the outputs one can investigate the ergotropy behavior in the wireless-like charging process of $QB$.

\section{Conclusion}\label{V}
In summary, we have investigated the dynamics of ergotropy for quantum batteries in common dissipative
bosonic environments. To this purpose, we have considered the time evolution of a two-qubit system mediated by a common environment with two approaches: non-Markovian dynamics and Markovian dynamics. In the first approach, we have shown the environment-mediated charging process (wireless-like charging), where the battery can be favorably charged when we are in a strong coupling regime and also provided an optical experimental setup to evaluate the amount of extractable work in this scenario. Furthermore, we have studied the second approach for two cases: (i) single-cell battery and (ii) two-cell battery. Our results show that an underdamped regime and/or  strong external fields can play essential roles in the optimal charging process of open quantum batteries in Markovian dynamics. Also, our models lighting the way to have stable and robust quantum batteries in the future. Moreover, we have found that in some scenarios, initial
quantum correlations between the charger and the battery or between the battery components may not be a useful resource for further extractable work.

\begin{acknowledgments}
This work has been supported by the University of Kurdistan. F. T. Tabesh and S. Salimi thank Vice Chancellorship of Research and Technology,  University of Kurdistan. We thank Alan C. Santos and Romain Bachelard for very useful discussions during the development of this work.
\end{acknowledgments}


\end{document}